%% file: Uplifting.tex
\documentclass[a4paper,12pt]{article}
\usepackage[T1]{fontenc}
\usepackage[utf8]{inputenc}
\usepackage[nosort]{cite}
\usepackage{catchfilebetweentags}
\usepackage{setspace}
\usepackage{amsmath,amssymb,amstext,amsthm,amssymb,amsfonts}
\usepackage{overpic}
\usepackage{float}
\usepackage{subcaption}
\usepackage{bbold}
\usepackage{bbm}
\usepackage{array}
\usepackage{lipsum}
\usepackage{enumitem}
\usepackage{cancel}
\usepackage{graphicx}
\usepackage{color}
\usepackage{slashed}
\usepackage{etoolbox}
\usepackage{hyperref}

\graphicspath{{figures/}}

\newcommand{\abs}[1]{\left|#1\right|}
\newcommand{\calO}{{\cal O}}
\newcommand{\calV}{{\cal V}}

\renewcommand{\Im}{\text{Im}\,}
\renewcommand{\Re}{\text{Re}\,}

\newcommand{\ap}{\alpha'}

\newcommand{\Pl}{\text{P}}

\newcommand{\eq}[1]{\begin{equation}#1\end{equation}}
\newcommand{\aeq}[1]{\begin{equation}\begin{aligned}#1\end{aligned}\end{equation}}
\newcommand{\ceq}[1]{\begin{equation}\begin{gathered}#1\end{gathered}\end{equation}}

\begin{document}
 \thispagestyle{empty}
 \begin{center}
 
  \vspace*{1.0cm}
  
  {\Large\bf 
   Winding Uplifts and the Challenges of\\[.3cm]
   Weak and Strong SUSY Breaking in AdS
  }
   
  \vspace*{1.0cm}
  
  {\large Arthur Hebecker and Sascha Leonhardt}\\[0.6cm]
  
  {\it
   Institute for Theoretical Physics, Heidelberg University,\\ Philosophenweg 19, 69120 Heidelberg, Germany\\[3mm]
   {\small\tt (\,a.hebecker, s.leonhardt~@thphys.uni-heidelberg.de\,)}
  }\\[1.5cm]
 \end{center}
 
 \begin{abstract}\normalsize
  We discuss the problem of metastable SUSY breaking in the landscape. While this is clearly crucial for the various de Sitter proposals, it is also interesting to consider the SUSY breaking challenge in the AdS context. For example, it could be that a stronger form of the non-SUSY AdS conjecture holds: It would forbid even metastable non-SUSY AdS in cases where the SUSY-breaking scale is parametrically above/below the AdS scale. At the technical level, the present paper proposes to break SUSY using the multi-cosine-shaped axion potentials which arise if a long winding trajectory of a `complex-structure axion' appears in the large-complex-structure limit of a Calabi-Yau orientifold. This has been studied in the context of `Winding Inflation', but the potential for SUSY breaking has not been fully explored. We discuss the application to uplifting LVS vacua, point out the challenges which one faces in the KKLT context, and consider the possibility of violating the non-SUSY AdS conjecture in the type-IIA setting of DGKT.
 \end{abstract}
 \newpage
 
 
 \section{Introduction and Summary}
 
 \subsection{Weak and Strong SUSY Breaking in the Landscape}
  
  Explicitly realizing de Sitter vacua in string theory is a long-standing challenge. The most popular approaches \cite{Kachru:2003aw, Balasubramanian:2005zx} start with an AdS solution with stabilized moduli and promote it to a de Sitter vacuum by a so-called `uplift'. In 4d supergravity language, such effects are classified as $F$-term or $D$-term uplifts (see e.g.~\cite{Saltman:2004sn, GomezReino:2006dk, Lebedev:2006qq, Dine:2006ii, Brummer:2006dg, Dudas:2006gr, Kallosh:2006dv, Abe:2006xp, Abe:2007yb, Achucarro:2007qa, Cicoli:2013cha, Cicoli:2015ylx, Bergshoeff:2015jxa,Retolaza:2015nvh,Argurio:2020dkg,Argurio:2020npm} and \cite{Burgess:2003ic, Villadoro:2005yq, Achucarro:2006zf, Dudas:2006vc, Haack:2006cy, Cremades:2007ig, Krippendorf:2009zza} respectively). All known models share a certain degree of complexity, which has lead to fundamental criticism \cite{Danielsson:2018ztv} and the proposal of corresponding no-go theorems \cite{Obied:2018sgi, Garg:2018reu, Ooguri:2018wrx}. If string theory really has a problem with de Sitter, one may wonder whether the SUSY-breaking uplift is its true source. 

  In particular, the anti-D3-brane uplift of KKLT \cite{Kachru:2003aw} tends to be uncomfortably high compared to the depth of the underlying AdS vacuum \cite{Carta:2019rhx}. It has been argued that a potentially fatal `singular-bulk problem' results \cite{Gao:2020xqh}.\footnote{
   We do not enter the interesting debate about the 10d description of KKLT (see e.g.~\cite{Moritz:2017xto, Gautason:2018gln, Hamada:2018qef, Kallosh:2019axr, Hamada:2019ack, Gautason:2019jwq, Carta:2019rhx}) since we believe that this is not going to invalidate the construction. We also note but do not discuss further the recently considered issues of throat-instabilities \cite{Bena:2018fqc, Blumenhagen:2019qcg, Randall:2019ent} and tadpole constraints \cite{Bena:2020xrh} (see however \cite{Crino:2020qwk}). Yet another line of attack is \cite{Sethi:2017phn}.
  }
  Combining these observations with a possible general unease about SUSY-breaking uplifts, one might suspect more concretely that {\it parametrically small} SUSY breaking is problematic in string theory compactifications. In this work, we will try to construct such small uplifts using the tuning power of the complex-structure landscape. We note that \cite{Palti:2020tsy} provides an alternative Swampland discussion of the SUSY breaking scale, relating it to the tower of light states. So far, we do not see an obvious relation to our work, but it might be interesting to return to this in the future.
  
  Before describing our approach, let us briefly consider existing and suggest some further Swampland constraints relevant in this context. First, the non-SUSY AdS conjecture states that no stable non-supersymmetric (4d) AdS exists in string theory \cite{Ooguri:2016pdq, Freivogel:2016qwc}. This is interesting for us since a small SUSY-breaking uplift on the basis of a SUSY AdS vacuum might provide a counterexample.

  A logically possible, much stronger conjecture would be one forbidding metastable non-SUSY AdS.\footnote{
   By metastable we mean that the local decay rate satisfies $\Gamma\ll R_{\rm AdS}^{-1}$. In spite of the fact that global metastable AdS decays instantaneously \cite{Horowitz:2007pr, Harlow:2010az}, a patch of such a metastable AdS might nevertheless exist, e.g.~as a cosmologically created bubble.
  }
  One may call this `Absolute non-SUSY AdS Conjecture'. In such a strong formulation this conjecture is in conflict with the constructions of \cite{Horowitz:2007pr, Narayan:2010em, Almuhairi:2011ws, Guarino:2020jwv, Guarino:2020flh}.\footnote{ 
   The recent discussions of \cite{Guarino:2020jwv,Guarino:2020flh} (based on \cite{Fischbacher:2010ec}) suggest that their solutions are stable. However, it is not clear to us to which extent non-perturbative instabilities can be excluded.
  }
  It then remains an interesting question whether some softened form of such a conjecture has a chance of being true. Most naively, one may think of flux compactifications where AdS and KK scale coincide, $R_{\rm AdS}\sim R_{\rm KK}$. If SUSY is broken by the compactification, one additionally expects $M_{\cancel{\rm SUSY}}  \sim R_{\rm KK}^{-1}$. It may then turn out to be difficult or impossible to escape the prediction that SUSY-breaking and AdS scale are related.\footnote{We 
  should remind the reader of the possibility of meta-stable AdS compactifications of non-SUSY string theories (the super-critical or the $O(16)^2$ heterotic string, see e.g.~\cite{Silverstein:2001xn, Maloney:2002rr, Abel:2015oxa} and refs.~therein). If such a compactification can be realized and if the AdS and string scale can be parametrically separated, our motivation for $M_{\cancel{\rm SUSY}} \sim R_{\rm KK}^{-1}$ fails.
 }
  
  Concretely, one might expect that metastable non-SUSY AdS with
  \linebreak 
  $M_{\cancel{\rm SUSY}} \gg R_{\rm AdS}^{-1}$ is forbidden. Let us call this the `Strong-SUSY-Breaking Conjecture'. If this conjecture were false and the Swampland dS conjecture true, then a natural place for observers like us to find themselves in would be universes with a small negative cosmological constant. In other words, the `Strong-SUSY-Breaking Conjecture' removes an anthropically interesting part of the multiverse.\footnote{
   While our world appears not to belong to this part, even this is not entirely certain \cite{Hardy:2019apu}.
  }
  
  By contrast, one can also consider a conjecture stating that metastable non-SUSY AdS with $M_{\cancel{\rm SUSY}}\ll R_{\rm AdS}^{-1}$ is forbidden. This could be called the `Weak-SUSY-Breaking Conjecture'. It is motivated by the difficulty, mentioned above, to make the anti-D3-uplift of KKLT as small as desired. We will see below that our proposal applied to the DGKT vacuum \cite{DeWolfe:2005uu} may provide a counterexample to the latter (`Weak') but not to the former (`Strong') SUSY-Breaking Conjecture.

  Clearly, all of the above is strongly affected if one takes the existence of the LVS AdS vacuum for a fact, even if this vacuum were only metastable. The `Strong-SUSY-Breaking Conjecture' then immediately falls and it is likely that, through an appropriate uplift (for example the one proposed in this paper), the dS conjecture also fails. If the Winding Uplift we suggest turns out not to work, the `Weak-SUSY-Breaking Conjecture' may coexist with LVS AdS vacua.
  
  We also note that a conjecture against the separation of KK and AdS scale \cite{Lust:2019zwm} (see also \cite{Gautason:2018gln}) would, if true, change much of the discussion above. We dismiss this for the purpose of this paper, expecting the KKLT AdS vacuum with the standard fine tuning of $W_0$ \cite{Denef:2004ze} (or its specific realization in \cite{Demirtas:2019sip}) to provide a counterexample.

 \subsection{Weak SUSY Breaking from a Winding Uplift}

  Let us now turn to the description of our technical work. We follow the original proposal by Saltman and Silverstein \cite{Saltman:2004sn} to realize an uplift by finding metastable local minima in the complex-structure scalar potential. We will use the tuning-power of the complex-structure-based flux landscape to ensure that the corresponding $F$-term is small and the SUSY breaking is controlled. Our method of choice are the multi-cosine-shaped axion potentials, in the spirit of \cite{Kim:2004rp}. Specifically, several cosine terms are superimposed if a long winding trajectory of a `complex-structure axion' appears in the large-complex-structure limit of a Calabi-Yau orientifold. This has been studied in the inflationary context as `Winding Inflation' \cite{Hebecker:2015rya} (see also \cite{Kobayashi:2015aaa, Bizet:2016paj, Blumenhagen:2016bfp, Wolf:2017wmu}), but the potential of this method for realizing weak SUSY breaking with long lifetimes has not been analyzed in detail. We will comment on the technically related uplifting suggestions of \cite{Hebecker:2014kva, Hebecker:2018yxs, March-Russell:2020lkq} in a moment.
  
  We consider type-IIB CY orientifold compactifications with the complex structure moduli $u$ and $v$ at large-complex-structure, i.e.~Im$\,u\,$, Im$\,v\gg1$. In this limit, the K\"ahler potential does not depend on $\Re u$ and $\Re v$, such that a shift symmetry arises. It is only broken by the flux superpotential. We may choose fluxes $M$ and $N$ in such a way that only the linear combination $M u + N v$ appears in the superpotential. As a result, one linear combination of $\Re u$ and $\Re v$ is left unstabilized. We parameterize this direction in field space by $\varphi \equiv \Re v$.
  
  The leading corrections to the large-complex-structure  expressions for K\"ahler and superpotential are of the form $\exp(iu)$ and $\exp(iv)$. Both terms depend on the unstabilized axion $\varphi$ and their magnitude is governed by the stabilized values of the saxions: $\exp\!\left(i u\right) \propto \exp\!\left(-\Im u_0-i N/M \, \varphi\right)$ and $\exp\!\left(i v\right) \propto \exp\!\left(-\Im v_0+i \varphi\right)$. We may tune the saxion values $\Im u_0$ and $\Im v_0$ in such a way that the two terms are comparable, suppressed by an expansion parameter $\epsilon \equiv \exp\!\left(-\Im u_0\right) \sim \exp\!\left(-\Im v_0\right) \ll 1$. Their relative magnitude is then measured by a parameter $\alpha \propto \exp\!\left(\Im u_0-\Im v_0\right) = \calO(1)$.
  
\begin{figure}[ht]
   \centering
   \def\svgwidth{.6\textwidth}
   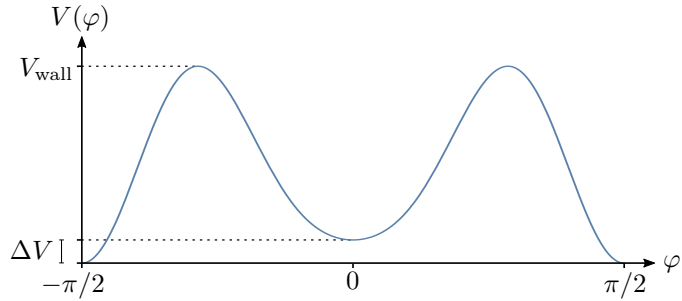
   \caption{The axion potential \eqref{eq:IntroPotential} for $N/M=3$. There is a minimum at $\varphi_* = 0$ with $\Delta V \equiv V(0) \propto g_s \epsilon^2 \gamma^2/\calV^2$, $\gamma \ll 1$, while the potential scales as $g_s \epsilon^2/\calV^2$ in general.}
   \label{fig:ZerosIntro}
  \end{figure}

  The resulting $F$-term scalar potential of the axion is, schematically, of the simple form
  \eq{\label{eq:IntroPotential}
   V(\varphi) = \frac{g_s}{\calV^2} \epsilon^2 \left[\cos(\varphi) - \alpha \cos(N/M \, \varphi)\right]^2\,.
  }
  Here, we introduced the string coupling $g_s$ and the CY volume $\calV$ as they usually appear in an $F$-term potential. Without loss of generality, we assume $M/N < 1$. In the regime $M^2/N^2 \lesssim \alpha \lesssim 1$, the $F$-term potential develops non-trivial local minima, e.g.~at some $\varphi=\varphi_*$ (cf.~Fig.~\ref{fig:ZerosIntro}). The value of the potential at the minimum takes the form $V(\varphi_*) \sim g_s \epsilon^2 \gamma^2/\calV^2$, where for $\varphi_*=0$ we define $\gamma \equiv 1- \alpha$. This can be tuned small if, as we will discuss in the paper, $\alpha$ is scanned sufficiently finely in the landscape such that $\gamma\ll 1$ can be realized. The height of the potential barrier separating the metastable from the global minimum is $V_\text{wall} \sim g_s \epsilon^2/\calV^2$. This remains sizable even at very small $\gamma$. Hence, the uplifting height $\Delta V$ and the height of the barrier $V_\text{wall}$ can be separated parametrically.
  
  Let us turn to possible applications of the mechanism just described in concrete settings: It is straightforward to apply it in the large volume scenario. Tuning the value of $\epsilon \gamma$ against the value of the LVS AdS cosmological constant, one may consistently uplift the vacuum to de Sitter. By contrast, uplifting a KKLT SUSY-AdS vacuum using the minimal setup just discussed is problematic. The difficulties one encounters are related to the smallness of the superpotential $W_0$, which is required for 10d supergravity control in KKLT. This smallness spoils the stabilization of the saxions as discussed above. The situation is not hopeless if one goes to the boundary of parametric control or involves more than two axions, but we have to leave a detailed study to future work. Finally, we consider supersymmetric AdS vacua in type IIA as studied by DGKT \cite{DeWolfe:2005uu}. This setting naturally gives rise to unstabilized axions in an otherwise fully supersymmetrically stabilized background. Only a single linear combination of RR axions is fixed. The superpotential resulting from non-perturbative corrections directly realizes our winding scenario with multiple axions \cite{Palti:2015xra}. We expect there is in general enough tuning power in this setting to find low-lying local minima protected by parametrically high barriers. While these uplifts are necessarily small, they may provide a way to turn DGKT solutions into stable non-SUSY AdS vacua.
  
  Before closing our introduction, let us comment in more detail on earlier related work. First, we note that `instantonic' terms have been used in various approaches to constructing de Sitter vacua, e.g., in racetrack or STU models \cite{Westphal:2006tn, deAlwis:2011dp, Rummel:2011cd, Cicoli:2012fh, Louis:2012nb, Blaback:2013qza, Rummel:2014raa, Braun:2015pza}. In some cases, these are one-step constructions, without the separation in AdS stabilization and uplift. A distinguishing feature of our approach is the complex-structure origin of the instantonic uplifting effect. This may allow a fully explicit implementation of our scenario, along the lines of the approach of \cite{Demirtas:2019sip} to constructing a small superpotential from the interplay of instantonic terms. We also note that combining a periodic and a linear potential for a complex-structure axion was suggested as an uplifting mechanism in \cite{Hebecker:2014kva}. Moreover, the interplay of different periodic terms in axion potentials has recently been discussed in other contexts: The `drifting monodromies' scenario in compactifications involving multi-throat systems \cite{Hebecker:2018yxs} gives rise to a superpotential of the same form as we use in this work. There, the particular no-scale form of the K\"ahler potential prevents a direct application to uplifting.\footnote{ 
  Due to the no-scale structure only the holomorphic part of the $F$-term enters the scalar potential. This leads to the same issues of implementing our mechanism as will be discussed in Sect.~\ref{sec:KKLT}. No-scale breaking effects may resolve this problem.
  }
  A closely related but slightly more speculative uplifting idea nevertheless arises in the setting of \cite{Hebecker:2018yxs}. The authors of \cite{March-Russell:2020lkq} give a pure IR argument on how the QCD pion potential at general $\theta$-angle generates a multi-cosine-shaped scalar potential which possesses non-zero minima. They discuss how this may naturally uplift the IR theory.
  
  Our paper is organized as follows. Sect.~\ref{sec:MainSection} presents the SUSY-breaking mechanism just discussed in detail, including a short introduction to the winding idea, a discussion of all relevant sub-leading corrections and an analysis of the axion potential induced. We describe how this may be used to uplift AdS vacua of various origins (LVS, KKLT and DGKT) in Sect.~\ref{sec:Uplifting}. Finally, we conclude in Sect.~\ref{sec:Conclusion}.

 \section{The Uplifting Potential}\label{sec:MainSection}

 \subsection{Winding Setup}\label{sec:Winding}

  Let us briefly introduce the winding scenario of \cite{Hebecker:2015rya}. It is formulated in a region of moduli space where two distinguished complex structure moduli $u$ and $v$ are `at large complex structure': $\Im u$, $\Im v \gg 1$. The K\"ahler potential is then approximately independent of $\Re u$ and $\Re v$, leading to a shift symmetry. Such approximate complex-structure shift  symmetries have been studied in the inflationary context in many papers, see e.g.~\cite{Hebecker:2014eua, Arends:2014qca, McAllister:2014mpa, Blumenhagen:2014nba, Hayashi:2014aua, Abe:2014xja, Garcia-Etxebarria:2014wla, Baume:2016psm, Hebecker:2018fln}.

  In the winding scenario, this shift symmetry is broken by the flux superpotential in such a way that a single axion-like field emerges. This field corresponds to a long, winding trajectory on the torus parameterized by $(\Re u\,, \Re v)$. Concretely, a flux choice is made such that the complex-structure superpotential and the full K\"ahler potential take the form
  \aeq{\label{eq:WindingSetup}
   W_\text{cs} &= \widetilde{W}_0(z) + f(z) (M u + N v) + W_\text{sub}(z,u,v) \,,\\
   K &= K_\calV - \ln(k(z,\overline{z},\Im u, \Im v)) + K_\text{sub}(z,\overline{z},u,\overline{u},v,\overline{v})\,.
  }
  Here, $K_\calV$ is the K\"ahler moduli K\"ahler potential. The variable $z$ represents the axio-dilaton together with all complex structure moduli, except for $u$ and $v$. Correspondingly, $-\ln(k)$ is the sum of axio-dilaton and complex-structure K\"ahler potential.\footnote{
   To be more explicit about the axio-dilaton $\tau$, one could replace $\{z\}\,\to\,\{\tau,z\}$, such that $- \ln\!\left(k(z,\overline{z},\Im u, \Im v)\right)\,\,\to\,\,-\ln\!\left(-2 \, \Im \tau\right) - \ln\!\left(\tilde{k}(z,\overline{z},\Im u,  \Im v)\right)$.
  } The expressions $W_\text{sub}$ and $K_\text{sub}$ stand for terms that are sub-leading w.r.t.
  \eq{ 
   W_0(z,u,v) \equiv \widetilde{W}_0(z) + f(z) (M u + N v) \,,
  \label{wws}}
  and
  \eq{
   K_0 \equiv K_\calV - \ln(k)\,.
  }
  These sub-leading terms are suppressed by factors $\exp(iu)$ or $\exp(iv)$ and arise as corrections to the periods $\int \Omega$ of the large-complex-structure geometry. We will specify these sub-leading terms in Sect.~\ref{sec:SubLeadingCorrections}. Crucially, by our flux choice $u$ and $v$ appear in $W_0$ only in the  linear combination $Mu+Nv$, where $M$ and $N$ are integer flux numbers. Note that we do not require large hierarchies in fluxes as will become clear later. Thereby, we avoid any potential issues arising in winding scenarios with large flux hierarchies \cite{Palti:2015xra,Hebecker:2018fln}.
  
  To analyze the leading-order $F$-terms $F_{0,i} \equiv (\partial_i + K_{0,i}) W_0$, it is convenient to change variables from $u,v$ to
  \eq{ \label{ppdef}
   \psi \equiv M u + N v \,, \quad \phi \equiv v \,.
  }  
  The $F$-term conditions then read
  \aeq{\label{eq:ZerothOrderFTerm}
   F_{0,z} &= (\partial_z K_0) W_0 + \partial_z \widetilde{W}_{0} + (\partial_z f) \psi = 0\,, \\
   F_{0,\psi} &= (\partial_\psi K_0) W_0 + f = 0 \,, \\
   F_{0,\phi} &= (\partial_\phi K_0) W_0 = 0 \,.
  }
  While the equations for $z$ and $\psi$ are in general complex, the equation for $\phi$ is real (up to an overall phase). The SUSY conditions $F_{0,i} = 0$ therefore fix
  \eq{ 
   z = z_0 \,, \quad \psi = \psi_0 \,, \quad \Im \phi = \Im \phi_0
  }
  while $\Re \phi$ remains unstabilized. As a result, the imaginary parts of the original fields $u$ and $v$ are also fixed,
  \eq{\label{eq:FTermSol}
   \Im u = \Im u_0 \,, \quad \Im v = \Im v_0 \,,
  }
  while only one linear combination of their real parts is stabilized. It will prove convenient to redefine the fields $z$, $\psi$ and $\phi$ according to
  \begin{equation}
  \label{fshi}
   z\,\to\,z_0+z\,\,,\qquad \psi\,\to\,\psi_0+\psi\,\,,\qquad \phi\,\to\,\phi_0+\phi\,,
  \end{equation}
  such that $z =\psi = \Im \phi = 0$ in the leading-order vacuum. (We set $\Re \phi_0 \equiv 0$, that is we do not shift the unstabilized field.) Since we do not apply this shift to $u$ and $v$, the relations \eqref{ppdef} must be appropriately corrected:
  \eq{\label{eq:ShiftedDef}
   \psi \equiv M u + N v - \psi_0 \,, \quad \phi \equiv v - \phi_0 \,.
  }

 \subsection{Sub-Leading Terms}\label{sec:SubLeadingCorrections}

  Sub-leading terms stabilize $\Re \phi$ and correct the vacuum values of the other fields.
  For the complex-structure superpotential, we have \cite{Hosono:1994av}\,\footnote{
   In fact, the underlying structure of corrections to the large-complex-structure expressions for periods holds both in the 3-fold and the 4-fold case. Thus, our discussion immediately applies to the more general F-theory setting. Note that prefactors like $A(z)$ may be viewed as arising from a full resummation of terms suppressed by $\exp(inz)$, $n\in \mathbb{N}$. 
  }
  \begin{eqnarray}
   \label{wl1}
   \hspace*{-.6cm}W_\text{cs} &\!\!\!\!\!=\!\!\!& W_0 + W_\text{sub} = W_0(z,u,v) + A(z) e^{i u} + B(z) e^{i v} + \ldots \\[.2cm]
   \label{wl2}
   &\!\!\!=\!\!\!& W_0(z,\psi) + A(z) e^{-\Im u_0} e^{i \Re \psi_0 / M}e^{i (\psi-N \phi)/M} + B(z) e^{-\Im v_0} e^{i \phi} + \ldots\\[.2cm]
   \label{wl3}
   &\!\!\!=\!\!\!& W_0(z,\psi) + \epsilon \, \left[ {\cal A}(z,\psi, \Im \phi)\, e^{-i N \varphi/M} + {\cal B} (z, \Im \phi)\, e^{i \varphi} \right] + \calO(\epsilon^2)\,,
  \end{eqnarray}
  where
  \eq{ 
   \epsilon \equiv e^{-\Im u_0} \quad \mbox{and} \quad \varphi \equiv \Re \phi\,.
  }
  In (\ref{wl2}), we have simply applied the field definitions from \eqref{eq:ShiftedDef}. Then, in (\ref{wl3}), we have absorbed all factors depending on fields that are stabilized in leading order in the two coefficients
  \aeq{ 
   {\cal A}(z,\psi, \Im \phi) &= A(z) e^{i\Re\psi_0/M} e^{i \psi/M + N \Im \phi / M} \,,\\
   {\cal B}(z,\Im \phi) &= B(z) e^{\Im u_0 - \Im v_0} e^{- \Im \phi} \,.
  }
  As a result, our expression for $W_\text{sub}$ in (\ref{wl3}) is manifestly a sum of two exponentials with different periodicities in the light axionic variable $\varphi$. While this is now somewhat hidden, $W_\text{sub}$ of course remains holomorphic in $\phi$.
  
  The large complex structure regime implies $\epsilon \ll 1$. Furthermore, we assume that by landscape-tuning 
  \eq{\label{eq:Tuning}
   \Im u_0 \simeq \Im v_0 \qquad \mbox{or} \qquad \abs{{\cal B}(0)/{\cal A}(0)} \propto e^{\Im u_0-\Im v_0} = \calO(1) \,,
  }
  such that the two sub-leading terms in  (\ref{wl3}) are comparable. We will specify the required $\calO(1)$-ratio more precisely below.
  
  We also add the relevant corrections to the K\"ahler potential:
  \aeq{\label{eq:OEpsKaehlerPot}
   K &= K_0 + K_\text{sub} = K_0(z,\overline{z},\Im u, \Im v) + K_\text{sub}(z,\overline{z},\phi,\overline{\phi},\psi,\overline{\psi}) \,,\\
   K_\text{sub} &= \left(\widetilde{A}(z,\overline{z},\Im u, \Im v) e^{i u} + \widetilde{B}(z,\overline{z},\Im u, \Im v) e^{i v} + \text{c.c.}\right) + \ldots \\
   &= \epsilon\, \left[ \widetilde{\cal A}(z,\overline{z},\Re \psi,\Im \psi, \Im \phi)\, e^{-i N \varphi/M} \right.\\
   &\qquad\quad \left. + \widetilde{\cal B} (z,\overline{z},\Im \psi, \Im \phi)\, e^{i \varphi} + \text{c.c.} \right] + \calO(\epsilon^2) \,.
  }
  Here $\widetilde{\cal A}$ and $\widetilde{\cal B}$ are defined similarly to ${\cal A}$ and ${\cal B}$. Note that our treatment of $K_\text{sub}$ as a sub-leading correction relies on the fact that the prefactors $\widetilde{A}$ and $\widetilde{B}$ depend on $\Im u$ and $\Im v$ only polynomially \cite{Hosono:1994av}.

 \subsection{The Axion Potential}\label{sec:AxionPotential}
  
  We now turn to the scalar potential induced by the sub-leading terms. Our analysis of the back-reaction on the leading-order solution simplifies the discussion presented in \cite{Hebecker:2015rya}. This will be useful for the generalization to small $W_0$ required later on.
  
  It will be convenient for the moment to include $\phi$ and $\psi$ in the set of complex structure moduli denoted by $z^i$, $i=1,\ldots,n$. So the index `$i$' now runs over {\it all} complex structure moduli and the axio-dilaton. We also shift all fields such that the leading-order vacuum is at $z^i=0$ for {\it all} $i$.

  Using the no-scale structure of the K\"ahler sector, the scalar potential takes the form\footnote{To 
  be precise, for this simple expression to be correct the 2-form-axion superfields (associated with non-zero $h^{1,1}_-$) have to be set to zero. In general, these fields appear in the K\"ahler moduli K\"ahler potential in combination with the axio-dilaton, which is one of our $z^i$. For non-zero 2-form axions, this affects the relevant $F_i$. Eventually, the potential is nevertheless independent of these 2-form axions because of their shift symmetry and an associated special no-scale cancellation \cite{Grimm:2004uq, Corvilain:2016kwe}.
  }
  \eq{\label{eq:fpot}
   V = e^K K^{i\overline{\jmath}} F_i \overline{F}_{\overline{\jmath}} \,,
  }
  where at zeroth order in $\epsilon$ we have $F_{0,i}=0$ at $z^i=0$ \eqref{eq:ZerothOrderFTerm}. At linear order in $\epsilon$, $F_i$ receives a correction $\delta F_i$ coming both from corrections to $K$ and $W$:
  \eq{\label{eq:DeltaF}
   \delta F_i = \partial_i K_0 W_\text{sub} + \partial_i K_\text{sub} W_0 + \partial_i W_\text{sub} \,.
  }
  But it would be too naive to simply replace $F_i$ in (\ref{eq:fpot}) by $\delta F_i$. The reason is that the $z^i$ back-react. This back-reaction is small, $z^i\sim \epsilon$, since $\delta F_i \sim \epsilon$.  We may thus Taylor expand in $z^i$ and evaluate (\ref{eq:fpot}) with the replacement
    \eq{\label{eq:fmod}
   F_i \quad \longrightarrow \quad F_{ij}z^j + F_{i\overline{\jmath}}\overline{z}^{\overline{\jmath}} + \delta F_i \,.
  }
  Here, $F_{ij}=\partial F_{0,i}/\partial z^j$ and similarly for $F_{i\overline{\jmath}}$. Since we are only interested in calculating $V$ at the order $\epsilon^2$, the $z^i$ dependence of $\exp(K)$ and $K^{i\overline{\jmath}}$ in \eqref{eq:fpot} may be disregarded. Moreover, the dependence of the small quantities $\delta F_i$ on the small parameters $z^i$ is irrelevant since it gives sub-leading terms in the $\epsilon$-expansion.
  
  In fact, the last statement comes with a crucial exception: Namely, our light axion $\varphi$ is now simply the real part of one of the $z^i$, and this field is not stabilized at leading order. Hence, in contrast to what was assumed about the generic $z^i$ above, this particular field excursions can take ${\cal O}(1)$ values. However, $\varphi$ appears in $\delta F_i$ and only there. Thus, our final result is (\ref{eq:fpot}) with the replacement \eqref{eq:fmod} and the extra prescription that $\delta F_i$ should be evaluated in full precision w.r.t.~$\varphi$ while keeping all other fields at their leading-order vacuum value, $z^i=0$.

  To proceed, let us view \eqref{eq:fpot} as the length squared of the complex vector $F_i$. At the expense of doubling the index range and appropriately redefining the metric, we may view this as the length squared of a real vector:
  \eq{\label{eq:DiagFTerms}
   V=G^{ab}f_a f_b \qquad \mbox{with} \quad  f_a=k_{ab}x^b+\delta f_a(x^1)\quad \mbox{and}
   \quad z^i=x^{2i-1}+ix^{2i} \,.
  }
  Here, we set $x^1\equiv \varphi$ such that the vector $k_{a1}$ vanishes by leading-order shift symmetry. The index range is $a,b=1,\ldots,2n$. The quantities $G^{ab}$, $f_a$, $\delta f_a$ and $k_{ab}$ follow from \eqref{eq:fpot} and \eqref{eq:fmod} by a simple rewriting in real and imaginary components.
  
  Our potential as a function of $x^1$ follows from \eqref{eq:DiagFTerms} by integrating out $x^2,\ldots, x^{2n}$, which is straightforward: The first term in $f_a$ generically takes values in a $(2n{-}1)$-dimensional subspace of the $\mathbb{R}^{2n}$ in which $f_a$ and $\delta f_a$ live. This is a result of $k_{a1}$ vanishing. Let us call the unit vector orthogonal to that `allowed' subspace $\hat{e}_a$. In this we use the inner product defined by $G^{ab}$. The vector $f_a$ can now be decomposed as the sum of its projection on $\hat{e}_a$ and its orthogonal projection. The potential is the sum of the squares of these two vectors:
  \begin{equation}\label{tte}
   V=\left| P_{\hat{e}}(f)\right|^2+\left| P_{\perp \hat{e}}(f)\right|^2\,.
  \end{equation}
  When minimizing in $x^2,\ldots, x^{2n}$ at fixed $x^1$ the vector $k_{ab}x^b$ will take the value $-\left(P_{\perp \hat{e}}(\delta f)\right)_a$ such that the second term in (\ref{tte}) vanishes. By contrast, the $\hat{e}_a$-subspace is not accessible to $k_{ab}x^b$, so one is simply left with the square of the projection of $\delta f_a$ on that subspace:
  \eq{\label{eq:BackReactedPot}
   V = \left| P_{\hat{e}}(f)\right|^2 = \left| P_{\hat{e}}(\delta f)\right|^2=\left(\hat{e}^a\,\delta f_a(x^1)\right)^2 \,.
  }
  The elements of $\hat{e}^a = G^{ab} \hat{e}_b$ may be calculated in terms of the vacuum values of $K_0$ and $W_0$. Given the form of $\delta F_i$ in \eqref{eq:DeltaF}, we see that the expression for the potential only contains sine and cosine terms in $\varphi$ with periodicity $2 \pi$ and $2\pi M/N$ as inherited from the complex exponentials in $W_\text{sub}$ and $K_\text{sub}$. We may hence choose to parameterize the potential as
  \eq{\label{eq:Potential}
   V(\varphi) = e^{K_0} \, \kappa \, \epsilon^2 \, \left[\cos(\varphi + \delta_1) - \alpha \cos(N\varphi/M + \delta_2)  \right]^2\,.
  }
  Here $\kappa$ and $\alpha$ are generically $\calO(1)$ coefficients and $\delta_{1,2}$ are phases arising in the transition from the complex to the real parameterization.\footnote{ 
   To be precise, the perturbations $\delta F_i(\varphi)$ in complex notation contain periodic terms $\propto e^{-i N \varphi/M}$ with coefficients $\partial_{z^i} K_0 {\cal A}$, $i N{\cal A}/M$,  $\partial_{z^i} {\cal A}$, $W_0 i N\widetilde{\cal A}/M$ and $W_0 \partial_{z^i} \widetilde{\cal A}$ as well as periodic terms $\propto e^{i \varphi}$ with coefficients $\partial_{z^i} K_0 {\cal B}$, $i {\cal B}$, $\partial_{z^i} {\cal B}$, $W_0 i \widetilde{\cal B}$ and $W_0 \partial_{z^i} \widetilde{\cal B}$. These coefficients are generically $\calO(1)$ when we are in the regime specified by \eqref{eq:Tuning}. Due to the projection on $\hat{e}$, the final result also depends on further (second) derivative terms of the leading-order K\"ahler and superpotential evaluated in $z^i=0$.
  } 
  It will be crucial that all the above coefficients and in particular $\alpha$ are tunable if a dense discretuum of vacua on the space parameterized by the $z^i$ exists. For example, $\alpha$ can be tuned using the ratio of exponentials of $u_0$ and $v_0$ in \eqref{eq:Tuning}.
  
  Finally, we express the K\"ahler potential $K_0$ through string coupling $g_s$ and CY volume $\calV$.   Absorbing numerical as well as $\tau$-independent terms in $k(0)$ in the prefactor $\kappa$, we arrive at
  \eq{\label{eq:AxionPotential} 
   V(\varphi) = \frac{g_s}{\calV^2} \, \kappa \, \epsilon^2 \left[\cos(\varphi + \delta_1) - \alpha \cos\!\left(\frac{N}{M} \varphi + \delta_2\right)\right]^2 \,.
  }
  Without loss of generality we assume a flux ratio $N/M > 1$.
  
\begin{figure}[ht]
   \centering
   \def\svgwidth{1.\textwidth}
   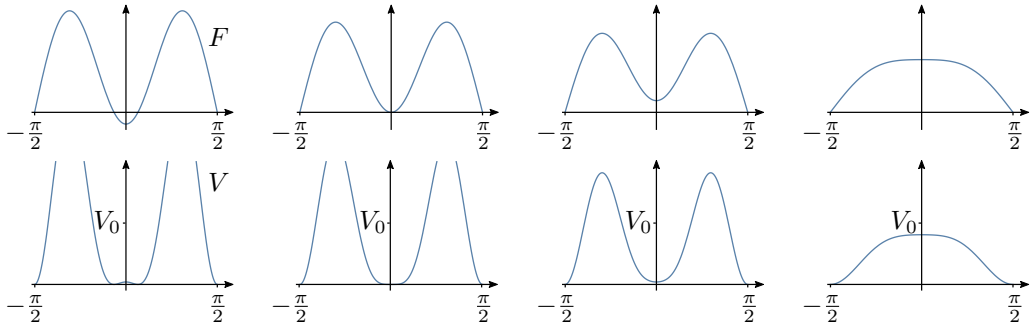
   \caption{The $F$-term $\propto [\cos(\varphi) - \alpha \cos(3 \varphi)]$ (upper panels) and corresponding scalar potential $V_0 [\cos(\varphi) - \alpha \cos(3 \varphi)]^2$ (lower panels) for $\alpha = 1.2\,,~1\,,~0.8\,,~0.1$ from left to right. By tuning $\alpha$ we find local minima at arbitrarily small positive value (third column). If $\alpha$ becomes too small the minima disappear (fourth column).}
   \label{fig:Zeros}
  \end{figure}

  To see that this potential possesses non-zero, local minima for tuned values of $\alpha$ and $\delta_i$, we consider the tuning $\delta_1 = \delta_2 = 0$. One easily finds that the potential has an extremum at $\varphi = 0$ with
  \eq{\label{eq:AxionPotMin}
   V(0) = \frac{g_s}{\calV^2} \, \kappa \, \epsilon^2 \, \gamma^2 \,, \quad V''(0) = 2 \frac{g_s}{\calV^2} \, \kappa \, \epsilon^2 \, \gamma \left[\frac{N^2}{M^2} \alpha - 1 \right] \,.
  }
  Here we have defined $\gamma \equiv (1-\alpha)$. The extremum is a minimum for $1 > \alpha > M^2/N^2$ and we may tune its potential value to be small by choosing $\gamma \ll 1$, see Fig.~\ref{fig:Zeros}. While also the second derivative becomes small, it goes to zero much more slowly: only linearly in $\gamma$.
  
  The decay constant and mass of $\varphi$ are given by
  \aeq{\label{eq:PhiMass}
   f_\varphi^2 &= K_{\phi\overline{\phi}} = \calO(1) \,,\\
   m_\varphi^2 &= V''(0)/ f_\varphi^2 = \calO(1) \, \frac{g_s}{\calV^2} \, \epsilon^2 \, \gamma \,\left[\frac{N^2}{M^2} \alpha - 1 \right].
  }
  Here we disregard the fact that, strictly speaking, $K_{\phi\overline{\phi}}$ is parametrically small by being suppressed by some power of $\Im u_0=\ln(1/\epsilon)$. Such minor effects are not essential in our context. Note that small flux ratios, $N/M-1 \ll 1$, suppress the mass of $\varphi$. For $\gamma \ll N^2/M^2-1$ the potential well around the minimum remains deep however.
  
  It is clear that an imperfect tuning, $\delta_i\simeq 0$, will not endanger our ability to adjust $\alpha$ and still realize a positive minimum at parametrically small potential value. There will also be qualitatively distinct, deep minima for different values of $\delta_i$. One could pursue their analytic study, but this does not appear necessary at present. For what follows, we will shift the field $\varphi$ such that the non-zero minimum remains at $\varphi=0$.
  
  With this we have arrived at one of our main technical results: We have provided an $F$-term uplift which, given enough tuning power in the complex structure landscape, can be extremely small and, in particular, small relative to the barrier protecting it from decay to supersymmetric minima.

Before closing this subsection and turning to generalizations and applications, let us summarize what our tuning requirements are precisely: First, we need a flux choice realizing the winding scenario, cf.~the second term on the r.h.~side of \eqref{wws}. For such a flux choice to exist, certain conditions have to be met by the integers which define the periods in the large complex structure limit (cf.~eq.~(6.6) of \cite{Hosono:1994av}). For example, terms of the type $z^i u$ and $z^i v$ have to be present. The relevant integers are mostly the triple-intersection numbers of the dual 3-fold. Second, the K\"ahler and superpotential have to be such that Im$\,u$ and Im$\,v$ are stabilized at an appropriately large value (at large complex structure). One might be concerned that this is a strong constraint since one linear combination of these moduli does not appear in $W$, where most of the tuning power resides. We note however that $K=K(u-\overline{u},v-\overline{v},z)$, if viewed as a function of Im$\,u$ and Im$\,v$, depends in detail on the values at which all the variables $z^i$ are stabilized. Hence the function $K(u-\overline{u},v-\overline{v})$ can be tuned through tuning the vacuum values of the $z^i$. We then expect to have the full tuning power of the complex-structure landscape at our disposal. Third, we need to tune the uplift potential, specifically the constants $\alpha$ and $\delta_{1,2}$ in \eqref{eq:AxionPotential}. Explicitly writing these constants in terms of $K$, $W$ and their derivatives would be cumbersome, but we have provided enough details above such that, in principle, such expressions can be derived for any given model. As explained previously for the tuning of Im$\,u$ and Im$\,v$, everything will depend on whether the flux landscape, viewed as a discrete set of points in the $z^i$-space, is dense enough. The seminal analysis of \cite{Denef:2004ze} and subsequent work appear to support this. Moreover, the very recent explicit analysis of \cite{Carta:2021sms} provides promising results concerning specifically the setup discussed in the present paper.

 \subsection{Winding in a Multi-Axion Field Space}\label{sec:GeneralCase}
  
  We may generalize to situations where multiple complex structure moduli $u^i$ ($i=0,\ldots,m$) are in the large-complex-structure limit and appear in $W$ only linearly at leading order. A possible superpotential that can arise in this case is
  \eq{ 
  \label{wmult}
   W_0(z,u^i) = \widetilde{W}_0(z) + f(z) \left(\sum_{i=0}^m N_i u^i \right) \,.
  }
  This is in fact similar to what happens generically in the type-IIA case to be discussed below.
  
  Now, defining
  \eq{ 
   \psi \equiv \sum_{i=0}^{m} N_i u^i  \quad \mbox{and} \quad \phi^i \equiv u^i \quad \mbox{for} ~~ i=1,\ldots,m \,,
  }
  we find the leading-order $F$-term conditions
  \eq{ 
   F_{0,\psi} = (\partial_\psi K_0) W_0 + f(z) = 0 \,, \qquad F_{0,\phi^i} = (\partial_{\phi^i} K_0) W_0 = 0\,.
  }
  By the same reasoning as in Sect.~\ref{sec:Winding} all imaginary parts are stabilized, $\Im u^i = \Im u^i_0$. By contrast, only one of the real parts is fixed, $\Re \psi = \Re \psi_0$. The remaining $m$ axions are massless at leading order. Shifting the fields $\psi$ and $\phi^i$ as in
  (\ref{fshi}), the corrected superpotential takes the form
  \aeq{\label{eq:MultipleAxions}
   W_\text{cs} &= W_0 + A_0 \, e^{-\Im u^0_0} \, e^{i \frac{\Re \psi_0}{N_0}} e^{-i \sum_{i=1}^{m} \left( \frac{N_i}{N_0} \phi^i \right)} + \sum_{i=1}^{m} A_i e^{-\Im u^i_0} \, e^{i \phi^i} + \ldots \\
   &\equiv W_0 + \epsilon \left[{\cal A}_0 \, e^{-i \sum_{i=1}^{m} \left( \frac{N_i}{N_0} \varphi^i \right)} + \sum_{i=1}^{m} {\cal A}_i \, e^{i \varphi^i} \right] + \calO(\epsilon^2)\,,
  }
  with the ${\cal A}_i$ defined as in Sect.~\ref{sec:SubLeadingCorrections}. The K\"ahler potential is of a similar form. As long as all coefficients $e^{-\Im u^i_0}$ ($i=0,\ldots,m$) are of the same order $\epsilon$, the resulting $F$-terms may be tuned to behave analogously to Fig.~\ref{fig:Zeros}. Of course, this now occurs over a higher-dimensional field space. 
  
  Another, maybe more interesting generalization arises if the superpotential of (\ref{wmult}) involves {\it different} $z$-de\-pendent prefactors \cite{Hebecker:2018fln}. We illustrate this using the 3-axion case by assuming a superpotential of the form
  \eq{ 
   W_0(z,u^i) = \widetilde{W}_0(z) + f(z) \left(\sum_{i=1}^3 M_i u^i \right) + g(z) \left(\sum_{i=1}^3 N_i u^i \right) \,.
  }
  A natural parameterization is now provided by
  \eq{ 
   \psi^1 \equiv \sum_{i=1}^{3} M_i u^i \,, \quad \psi^2 \equiv \sum_{i=1}^{3} N_i u^i \,, \quad \phi \equiv u^3 \,.
  }
  Crucially, only a single axion,  $\varphi \equiv \Re \phi$, remains unstabilized \cite{Hebecker:2018fln}. Repeating the exercise of adding sub-leading corrections results in a superpotential
  \eq{\label{eq:MultipleAxions2}
   W_\text{cs} = W_0 + \epsilon \left[{\cal A}_1 \, e^{-i \frac{M_3 N_2 - M_2 N_3}{M_1 N_2 - M_2 N_1} \varphi} + {\cal A}_2 \, e^{-i \frac{M_3 N_1 - M_1 N_3}{M_2 N_1 - M_1 N_2} \varphi} + {\cal A}_3 \, e^{i \varphi} \right] + \calO(\epsilon^2)\,.
  }
  This may be generalized further to $m\!+\!1$ fields $u^i$ appearing in $m$ linear combinations in the leading-order superpotential. After integrating out heavy fields, the resulting $F$-term potential will be a periodic function of a single axion, but with many tunable parameters ${\cal A}_i$. We will discuss the possible importance of this extra tuning freedom in the application to KKLT below.
  
  Finally, we may combine the previous generalizations by considering $m+1$ axions appearing in the superpotential in $n$ linear combinations (with $1\leq n \leq m$). The result will be a sub-leading potential for $m-n+1$ axions.

 \section{Uplifting AdS Vacua}\label{sec:Uplifting}

  In this section, we finally turn to our main goal: the uplifting of known AdS vacua to higher-lying AdS and dS solutions. Our interest is in testing conjectures against non-SUSY AdS and dS models \cite{Ooguri:2016pdq, Freivogel:2016qwc, Garg:2018reu, Ooguri:2018wrx} .
  
  Before turning to explicit scenarios, we want to highlight the key feature of our complex-structure $F$-term scalar potential \eqref{eq:AxionPotential}: Let $\Delta V = V_{f} - V_{t} \propto g_s \epsilon^2 \gamma^2/\calV^2$ be the difference between the non-zero false-vacuum value $V_f$ in the minimum \eqref{eq:AxionPotMin} and the true, global minimum at $V_t = 0$. This $\Delta V$ is tunable via the complex structure, which we characterize by the potentially very small parameter $\gamma$. When including non-trivial effects of K\"ahler moduli, both $V_f$ and $V_t$ will change. Nevertheless, we expect that $\Delta V$ will remain small if $\gamma$ was tuned to a tiny value.  By contrast, the potential barrier between $V_f$ and $V_t$ as well as the field distance $\Delta \varphi$ between vacua is independent of $\gamma$. This allows for a high degree of stability against false vacuum decay via Coleman-de Luccia bubble nucleation \cite{Coleman:1980aw}.\footnote{
   We have not investigated the possibility that decays to bubbles of nothing \cite{Witten:1981gj} affect our non-SUSY vacuum \cite{GarciaEtxebarria:2020xsr}. While we do not see how this would happen in our setting, a more careful study may be warranted.
  }
  
  We now turn to explicit scenarios of uplifting.

 \subsection{Large Volume Scenario}
  
  The axion potential \eqref{eq:AxionPotential} was derived assuming $\abs{W_0} = \calO(1)$. This is consistent with the large volume scenario for K\"ahler moduli stabilization  \cite{Balasubramanian:2005zx}. Thus, we may straightforwardly apply our method. Of course, many studies dedicated to uplifting the LVS exist \cite{Cremades:2007ig, Krippendorf:2009zza, Rummel:2011cd, Gallego:2011jm, Cicoli:2012vw, Cicoli:2013cha, Rummel:2014raa, Cicoli:2015ylx, Gallego:2017dvd}. Moreover, the warped anti-D3-brane uplift of KKLT \cite{Kachru:2003aw} is also applicable to the LVS. However, our approach stands out because it is truly minimal: It only uses ingredients which are already present in the LVS AdS vacuum, employing the tuning power of the flux-landscape rather then extra features like matter sectors or throats.
  
  We want to highlight \cite{Gallego:2017dvd} which is based on \cite{Kallosh:2014oja,Marsh:2014nla}. There, a general uplift mechanism in the continuous flux approximation is presented: In the LVS fluxes are chosen such that the complex-structure $F$-terms do no longer vanish, $F_i = \epsilon W_0 f_i \neq 0$. Here, $f_i$ is a unit-vector in the complex-structure moduli space and $\epsilon \ll 1$ may be tuned such that the corresponding scalar potential $V = g_s \epsilon^2 \abs{W_0}^2/\calV^2$ uplifts the LVS AdS vacuum. It is shown that the assumed structure of $F_i$ necessarily implies that one real direction of the complex-structure moduli space remains light. While the idea of uplifting via small SUSY-breaking $F$-terms at the cost of an additional light field is as in our mechanism, we do not require the assumption of continuous fluxes to achieve the required tuning.

  The LVS requires (at least) two K\"ahler moduli $T_b$ and $T_s$. Since the stabilization will eventually realize a hierarchy of the corresponding four-cycle volumes, $\tau_b \equiv \Re T_b \gg \Re T_s \equiv \tau_s$, only the dominant non-perturbative correction $W_\text{np} \propto e^{- a_s T_s}$ needs to be considered. In addition, the leading perturbative $\ap$-effect \cite{Becker:2002nn} is essential. It corrects the K\"ahler potential according to $-2\ln({\cal V}) \to -2\ln({\cal V+\xi})$. Here, ${\cal V}=\tau_b^{3/2}-c\tau_s^{3/2}$ and, for our purposes, $\xi\sim g_s^{-3/2}$ is a constant that can be tuned by choosing the stabilized value of $g_s$. The resulting $F$-term scalar potential has a minimum in $\tau_s$, which we hence integrate out. One is then left with a potential for $\calV \approx \tau_b^{3/2}$, which we parameterize as
  \eq{\label{eq:VPotential}
   V_\text{LVS}(\calV) = A \, \frac{g_s \xi \abs{W_0}^2}{\calV^3} - B \, \frac{g_s \abs{W_0}^2}{\calV^3} \ln(\calV)^{3/2} + \calO\left(\frac{g_s \ln(\calV)^{1/2}}{\calV^3}\right)\,,
  }
  where $A$ and $B$ are positive $\calO(1)$-coefficients. This potential has an AdS minimum at 
  \begin{equation}\label{eq:LVSMinimum}
   V_{\rm AdS} = - \calO(1) \frac{g_s \abs{W_0}^2 \ln(\calV)^{1/2}}{{\cal V}^3} \,,
  \end{equation}
  with ${\cal V}\sim \exp(a_s\tau_s)$ and $\tau_s\sim \xi^{2/3}$. There is a mass hierarchy $m_{\tau_b} \ll m_{\tau_s}$, justifying a posteriori the procedure just described. To be precise, from \eqref{eq:VPotential} and the K\"ahler potential we find
  \eq{\label{eq:VMass}
   m_{\tau_b} = \calO(1) \frac{g_s \abs{W_0}^2 \ln(\calV)^{1/2}}{\calV^3} \,.
  }
  
  The above assumed a vanishing complex-structure $F$-term scalar potential. Our main point is in relaxing this assumption and adding the complex-structure uplift of Sect.~\ref{sec:AxionPotential}:
  \eq{\label{eq:LVS+Uplift}
   V(\calV,\varphi) = V_\text{LVS}(\calV) + \frac{g_s \kappa \epsilon^2}{\calV^2} f(\varphi) \,.
  }
  Here $f(\varphi)$ is a non-negative periodic function (cf.~\eqref{eq:AxionPotential}) with supersymmetric minima at $f=0$ and a SUSY-breaking minimum at $f(0)=\gamma^2$. We choose to tune
  \eq{\label{eq:LVSTuning}
   \kappa \epsilon^2 \gamma^2 = \calO(1) \frac{ \abs{W_0}^2 \ln(\calV)^{1/2} }{\calV} \,,
  }  
  where $\calV$ is the LVS value of the volume. This tuning corresponds to an uplift to a Minkowski, shallow AdS or low-lying dS vacuum. One easily checks that such an uplift does not destabilize ${\cal V}$ and that a mass hierarchy $m_{\tau_s} \gg m_{\tau_b}, m_\varphi$ is obeyed. We hence do not need to reconsider the step of integrating out $\tau_s$.  
   
  The potential \eqref{eq:LVS+Uplift} describes a 2-field model with a mass ratio $m_{\tau_b}/m_\varphi \sim \sqrt{\gamma}$ \, (cf.~\eqref{eq:PhiMass} and \eqref{eq:VMass}). The condition \eqref{eq:LVSTuning} may be realized with either $\gamma \ll 1$ or $\gamma= \calO(1)$. In the first case, the complex-structure axion is lighter than the volume modulus, in the second case they have similar masses.
  
  Note that, while the possibility of an extremely small uplift due to a tuning $\gamma\ll 1$ is a distinguishing feature of our approach, this is not required for the LVS uplift. Indeed, in the present case, a value $\gamma = \calO(1)$ does not spoil the longevity of the vacuum: While the potential barrier between true AdS and false dS vacuum is not parametrically high compared to $\Delta V$, the distance $\Delta \varphi$ between vacua remains sufficiently large. In summary, if the tuning we assumed can be explicitly realized, our mechanism could challenge the dS conjecture \cite{Garg:2018reu,Ooguri:2018wrx}.
  
  The analysis of this section is built on the very reasonable expectation that the SUSY-breaking AdS vacuum of the LVS approach is long-lived. It is in fact even possible that this vacuum is stable and therefore in conflict with the non-SUSY AdS conjecture \cite{deAlwis:2013gka, Conlon:2018vov}. In either case, i.e.~both for a long-lived and for a stable LVS AdS vacuum, the possibility of our small uplift to a higher-lying AdS does not improve its usefulness as a counterexample to the non-SUSY AdS conjecture. Therefore, we do not pursue this further.
  
  Before closing the present section, we should comment on  a geometric consistency issue related to the large-complex-structure limit. In this limit, the CY can be thought of as a $T^3$-fibration over an $S^3$ with the fiber volume becoming singular \cite{Strominger:1996it, Kontsevich:2000yf, zbMATH01782659}.  Assigning a typical radius $R$ to the torus-fiber and a radius $L$ to the base three-sphere, the imaginary parts of the complex-structure moduli (in the conventions of this paper) scale as $L/R$ in the limit $L/R\to \infty$ (cf.~\cite{Arends:2014qca}). Since CY 2-cycles scale as $L\cdot R$ and since $R>1$ in string units is required for supergravity control, a lower bound on the volume is obtained, $\calV \gtrsim L^3$. To be precise, the moduli we assumed to be at large complex structure are $u$ and $v$. We then have Im$\,u \sim\,$Im$\,v\sim L/R$ and find the condition 
  \eq{\label{eq:SYZPicture}
   \Re T_b \sim {\cal V}^{2/3} \,\,\gtrsim\,\, (\Im u)^2 \sim (\Im v)^2\,.
  }
  This is easily consistent with the tuning requirement \eqref{eq:LVSTuning}. Indeed, since $\epsilon = \exp(-$Im$\,u)$, our volume is {\it exponentially} large in Im$\,u$. It becomes even larger if $\gamma\ll 1$ or $|W_0|\gg1$. For the small 4-cycle, a condition analogous to \eqref{eq:SYZPicture} is not obeyed so easily since Re$\,T_s$ grows only like $\ln{\cal V}$. It may still hold if $\gamma$ is tuned sufficiently small. But, most importantly, we expect that we do not even need to implement a geometric consistency condition like Re$\,T_s\gg (L/R)^2$ because our large-complex-structure limit is only partial: All the complex structure moduli $z^i$ apart from $u$ and $v$ are not required to be at large complex structure. Hence, we may focus on geometries where the blow-up cycle governed by $T_s$ is geometrically separated from the specific shrinking 3-cycles related to $u$ and $v$. In such geometries, we expect that supergravity control is straightforwardly compatible with the partial large-complex-structure limit we require.
  
 \subsection{The KKLT AdS Vacuum}\label{sec:KKLT}
  
  So far we considered non-tuned flux-superpotentials: $\abs{W_\text{cs}}\simeq \abs{W_0}\gtrsim {\cal O}(1)$. Let us now turn to uplifting the supersymmetric type-IIB KKLT vacuum \cite{Kachru:2003aw} which relies on the tuning $\abs{W_\text{cs}}\ll 1$. Note that our convention differs from \cite{Kachru:2003aw} in that we denote the complex-structure superpotential by $W_\text{cs}$ and reserve the symbol $W_0$ for the contribution that is leading at large complex structure: $W_\text{cs} \equiv W_0 + W_\text{sub}$. After including the K\"ahler modulus superpotential\footnote{ 
   We disregard the dependence of $A_K$ on the complex structure moduli $z^i$ and $u/v$ since the former are stabilized at a high scale and the latter enter only in a subdominant way, i.e.~as $e^{iu/v}$ with Im$\,u/v\gg 1$. This last statement follows from analyticity and periodicity in the real direction, as is briefly mentioned in \cite{Demirtas:2019sip} and may also be established rigorously \cite{hms}.
  }
  \begin{equation}
   W_\text{np}(T) = A_K \, e^{-a T}  \,,
  \end{equation}
  the modulus $T$ is stabilized at a value where $e^{-a\, \Re T} \sim \abs{W_\text{cs}}$.
  
  We notice that there is a problem in implementing the simple mechanism as described in Sect.~\ref{sec:AxionPotential}: If we consider a CY at the large-complex-structure point, we are subject to \eqref{eq:SYZPicture}. For the realization of the KKLT scenario in type IIB, this implies
  \eq{\label{eq:GeometricConsistency}
   \abs{W_\text{cs}} = \abs{W_0 + W_\text{sub}} \sim e^{-a \Re T} \ll e^{-\Im u} \sim \epsilon \,.
  }
  Therefore, the perturbative analysis of Sect.~\ref{sec:MainSection} relying on $\epsilon \ll \abs{W_0}$ breaks down.
 
  There are two possibilities to proceed: First, one may try to find a flux choice such that $|W_0+W_{\rm sub}|\ll\epsilon$. Since generically $W_{\rm sub}\sim \epsilon$, this requires a $W_0$ of the same magnitude. This means that our calculational approach, which treats $W_{\rm sub}$ as a small correction, is at the boundary of control and becomes unreliable. Nevertheless, one may hope that there are concrete models in which the qualitative features of our uplifting method survive.
  
  Second, one may try to implement a hierarchy $|W_{\rm sub}|\ll |W_0|\ll \epsilon$. This requires a fine-tuned cancellation between ${\cal O}(\epsilon)$-terms within $W_{\rm sub}$. The advantage is that one can hope to maintain the method of treating $W_{\rm sub}$ as a small correction to $W_0$.
  
  In either case, the whole superpotential $W=W_\text{cs}(\phi)+W_\text{np}(T)$ is small, which implies that not just the real component but the whole superfield $\phi$ remains light. We expect a consistent supergravity description to exist for the light moduli $\phi$ and $T$ with
  \eq{ 
   W = W_0 + W_\text{sub}(\phi) + W_\text{np}(T) \,, \quad K = -3 \ln(\Re T) - \ln(k(\Im \phi))
  }
  and
  \eq{\label{eq:PhiTPotential}
   V = e^K \left[K^{T\overline{T}} \abs{D_T W}^2 + K^{\phi\overline{\phi}} \abs{D_\phi W}^2 - 3 \abs{W}^2 \right] \,.
  }
  The lightness of $\Im \phi$ follows from the fact that it enters $V$ only via the small superpotential $W_\text{sub}$ or through the prefactors $K_\phi$, $K^{\phi\overline{\phi}}$ and $e^K$, which multiply small quantities. Note also that corrections to $K$ are not relevant for sufficiently small $W$.
  
  We proceed on the basis of the supersymmetric minimum at $D_T W = 0$ and $D_\phi W = 0$. Given our assumptions, this minimum is characterized by \cite{Kachru:2003aw}
  \eq{
   \abs{W_\text{cs}} \sim \abs{A_K} \, e^{-a\, \Re T} \,, \quad V_\text{AdS} \sim - \abs{W_\text{cs}}^2\,.
  }
  Following our previously defined strategy, we can now check the $F$-term potential of $\phi$ for nearby minima with $D_\phi W\neq0$. Such minima may lead to higher-lying AdS vacua of \eqref{eq:PhiTPotential} or even to metastable de Sitter. For this, we consider
  \eq{\label{eq:HolomorphicFTerm}
   D_\phi W = K_\phi(\Im \phi) \, W(\phi,T) + \partial_\phi W_{\rm sub}(\phi) \,.
  }
  If we choose the first of the two tuning options described above, where $W_{\rm sub}\sim \epsilon$ and $|W_0+W_{\rm sub}|\ll\epsilon$, a fundamental obstacle arises. Namely, in this regime generically $\abs{W} \ll \abs{\partial_\phi W_{\rm sub}(\phi)}$, such that the first term on the r.h.~side of \eqref{eq:HolomorphicFTerm} is negligible. But the second term is holomorphic in $\phi$. Hence, by the minimum modulus principle, its absolute value cannot have a non-zero local minimum. To overcome this obstacle, a further tuning is required: We need to ensure $\abs{\partial_\phi W_{\rm sub}(\phi)}\ll |W_{\rm sub}(\phi)|$ at the relevant point in $\phi$-space. Then one may hope that an interplay of the non-holomorphic first and the holomorphic second term on the r.h.~side of \eqref{eq:HolomorphicFTerm} produces the desired local minimum. 

  If we choose the second of the two tuning options described earlier, $|W_{\rm sub}|\ll |W_0|\ll\epsilon$, then it would naively appear that \eqref{eq:MultipleAxions2} can produce a non-trivial local minimum of the $F$-term potential along the lines of Sect.~\ref{sec:MainSection}. However, things are not that simple: The extraordinary smallness of $|W_{\rm sub}|$ comes from a compensation between different terms (e.g.~the two exponentials in \eqref{wl3}), and this cancellation does in general not extend to $\partial_\phi W_{\rm sub}$. A further tuning for small $\partial_\phi W_{\rm sub}$ is required. Still, the present tuning option may be advantageous since the first term on the r.h.~side of \eqref{eq:HolomorphicFTerm} simplifies: $K_\phi(\Im \phi) \, W(\phi,T)\simeq K_\phi(\Im \phi) \, (W_0+W_{\rm np}(T))$. This might allow us, similar to the scenario of Sect.~\ref{sec:MainSection}, to stabilize $\Im \phi$ independently of $\Re \phi$.

  Our preliminary investigation suggests that, employing either of the tuning options as discussed in the last two paragraphs, it is not straightforward to uplift KKLT. We expect that a sufficient amount of tuning freedom becomes available only if one has three or more exponential terms at one's disposal, cf.~\eqref{eq:MultipleAxions2}. Moreover, even after successfully engineering an $F$-term with a $(\Re \phi)$-dependence as in Fig.~\ref{fig:ZerosIntro}, one is not yet finished. Namely, since the whole superfield $\phi$ is very light one must check the non-trivial additional requirement that the full potential $e^K \! \left(K^{\phi\overline{\phi}}\abs{D_\phi W}^2-3\abs{W}^2\right)$ has a local $\phi$-minimum. Analyzing these problems is beyond the scope of this work. It would lead us too far away from our basic goal of highlighting the immediate applications of the winding potential \eqref{eq:AxionPotential} to the challenge of uplifting.

 \subsection{DGKT-type Vacua}
  
  While we derived the winding-uplift potential with a type-IIB compactification in mind, it is also possible and interesting to implement it in a type-IIA Calabi-Yau orientifold with fluxes. Specifically, we will consider DGKT vacua \cite{DeWolfe:2005uu}\footnote{ 
   Recent work includes generalizations \cite{Marchesano:2019hfb,Grimm:2019ixq} as well as checks of consistency \cite{Junghans:2020acz,Marchesano:2020qvg,Baines:2020dmu}, as triggered in particular by the AdS distance (or scale separation) conjecture \cite{Lust:2019zwm}.
  },
  following the notation of \cite{Palti:2008mg,Palti:2015xra}. In type IIA, the K\"ahler moduli $T^i$ as well as the axio-dilaton $S = s+i \sigma$ and the complex structure moduli $U_\lambda = u_\lambda + i \nu_\lambda$ appear in the perturbative flux superpotential: 
  \begin{equation}
   W_\text{flux} = W_\text{K}(T^i) + W_\text{cs}(S,U_\lambda)\,.
  \end{equation}
  Here $W_{\rm cs}$ combines the dilaton and complex-structure contributions, both involving 3-cycle data. It takes the explicit form
  \begin{equation} \label{csw}
     W_\text{cs}(S,U_\lambda) = -i h_0 S - i q^\lambda U_\lambda \,,
  \end{equation}
  where $h_0$ and $q^\lambda$ are independent $H_3$-flux numbers. Concerning $W_K$, it is sufficient to note that it involves terms up to cubic order in the 2-cycle variables $T^i$ (not to be confused with the 4-cycle K\"ahler moduli of type IIB, which we denoted by $T$ or $T_b, T_s$). The fluxes in $W_K$ come from Ramond-Ramond 0-, 2-, 4- and 6-form field strengths.
  
  The K\"ahler potential is given by
  \eq{ 
   K = -\ln(8 \calV) -\ln(S+\overline{S}) - 2 \ln( \calV') \,,
  }
  where $\calV(\Im T^i)$ is the type-IIA CY volume. The quantity ${\cal V}'$, which one may call the dual volume, is a function of the complex structure moduli $U_\lambda$. At large complex structure, it is defined implicitly by 
  \eq{ 
   \calV' \equiv \frac{d_{\lambda \rho \sigma}}{6} v^\lambda v^\rho v^\sigma \qquad \mbox{with} \qquad
   u_\lambda = \partial_{v^\lambda} \calV' \,.   
  }
  One may think of the $v^\lambda$ as 2-cycle volumes of the mirror dual type-IIB compactification \cite{Grimm:2004ua}, with $u_\lambda=\,\,$Im$\,U_\lambda$ characterizing the corresponding type-IIB 4-cycles. The $d_{\lambda \rho \sigma}$ are triple intersection numbers of the mirror Calabi-Yau. We see that $\calV'$ is a homogeneous function of degree $\frac{3}{2}$ in $u^\lambda$.
  
  The vanishing-$F$-term conditions for $S$ and $U_\lambda$ read
  \begin{eqnarray}
   \label{eq:IIACS1}
   2 h_0 s &=& -\Im W \,,  \\
    \label{eq:IIACS2}   
    K_{u_\lambda} &=& -\frac{q^\lambda}{h_0 s} \,,  \\
    \label{eq:IIACS3}   
    h_0 \sigma + q^\lambda \nu_\lambda &=& -\Re W_\text{K} \,.
  \end{eqnarray}
  We do not display the corresponding SUSY equations for the K\"ahler moduli. Suffice it to say that the volume $\calV$ may be considered a free parameter as it depends on the unconstrained 4-form fluxes appearing in $W_K$. In particular, one may go to large volume, where the following scaling behavior is found:
  \ceq{\label{eq:DGKTVacuum}
   \abs{W_0} \sim \Im W_\text{K} \sim \calV \quad, \qquad e^K \sim \calV^{-5} \\[.2cm]
   \Rightarrow \quad V_\text{AdS} \sim - e^K \abs{W_0}^2 \sim - \calV^{-3} \,.
  }
  
  We conclude two important facts about flux-stabilized type-IIA solutions and specifically \eqref{eq:IIACS1}-\eqref{eq:IIACS3}: First, all the real parts $u_\lambda$ of complex structure moduli  and the dilaton field $s$ are stabilized by fluxes. In particular, the ratios $\partial_{u_\lambda} \calV'/\partial_{u_\rho} \calV'$ are determined by $H_3$-flux ratios $q^\lambda/q^\rho$. The overall scale is set by $s$, which is in turn fixed by \eqref{eq:IIACS1}. Taking into account the in general rather complicated functional relation between the variables $\partial_{u_\lambda} \calV'$ and $u_\lambda$, one may expect a flux discretuum like in type IIB \cite{Denef:2004ze}. By this we mean that the lattice of allowed flux choices translates in a sufficiently dense and random set of points on the field space parameterized by $s$ and the $u_\lambda$. Note that the 3-form fluxes $h_0$ and $q^\lambda$ are subject to the $h^{2,1}_+ +1$ tadpole cancellation conditions $m_0 h_0 + Q^0_\text{D6} = 0$, $m_0 q^\lambda + Q^\lambda_\text{D6} = 0$, where $m_0$ is the Romans mass \cite{DeWolfe:2005uu, Palti:2015xra}. Thus, it needs to be checked whether in a given model the flux discretuum is dense enough for our purposes.

  Second, according to \eqref{eq:IIACS3} only a single linear combination of imaginary parts (the axions $\sigma$ and $\nu_\lambda$) is fixed. This ensures that a set of axions stays light, such that the winding proposal of \cite{Hebecker:2015rya} is automatically part of the DGKT setting \cite{Palti:2015xra}.
  
  Following \cite{Palti:2015xra}, we now include non-perturbative corrections to (\ref{csw}):
  \begin{equation}
   W_{\rm cs}\quad\to\quad W_{\rm cs} = W_0 + \sum_I A_I \exp\left(-a_0^I S - \sum_\lambda a_\lambda^I U_\lambda\right)\,.
  \end{equation}
  Here $I$ runs over all E2-instantons and the coefficients $a^I_0$/$a^I_\lambda$ specify the cycles wrapped by instanton $I$. Assuming that we are at large $s$ and $u_\lambda$ and that the dominant instantons contribute, this simplifies to 
  \begin{equation}
   W = W_0 + A_0\,  e^{-s}\, e^{-i\sigma} + \sum_{\lambda} A_\lambda\, e^{-u_\lambda} \, e^{-i\nu_\lambda} + \ldots
  \end{equation}
  We see that, including also the constraint of (\ref{eq:IIACS3}), this takes exactly the form of the superpotential \eqref{eq:MultipleAxions} from our previous type-IIB analysis. The following identifications make this explicit:
  \aeq{ 
   u_\lambda \quad &\longleftrightarrow \quad \Im u^i_0 \qquad \mbox{for}\quad i=\lambda=1,\ldots,m \\
   s \quad &\longleftrightarrow \quad \Im u^0_0 \\
   \nu_\lambda \quad &\longleftrightarrow \quad \varphi^i \qquad\quad~ \mbox{for}\quad i=\lambda=1,\ldots,m \\
   \sigma = -\Re W_\text{K}- \sum_{\lambda =1}^{m} \frac{q^\lambda}{h_0} \nu_\lambda \quad &\longleftrightarrow \quad \Re u^0 = \frac{\Re \psi_0}{N_0} - \sum_{i=1}^{m} \frac{N_i}{N_0} \varphi^i\,.
  }
  Thus, we may now think in terms of the multi-axion potential as displayed in the last line of \eqref{eq:MultipleAxions}  and discussed at length in Sect.~\ref{sec:MainSection}. A key point for our uplifting application was the smallness of $\epsilon$ and the tunability of the coefficients ${\cal A}_0$ and ${\cal A}_i$. Both is ensured if, as discussed above, an appropriate discretuum in the field space of $s$ and $u_\lambda$ exists.
 
  A difference to our type-IIB analysis is that here we automatically have many light axions. In Sect.~\ref{sec:MainSection}, a special flux choice was needed to keep one complex-structure axion light. Having several of them required more assumptions. We expect that there is nothing wrong with realizing our winding uplift in the type-IIA case in a multi-axion situation. But, as discussed in \cite{Palti:2015xra}, it is also easy to return to the single-axion case analyzed in detail before:
 
  First, we can choose a CY with a small number of complex structure moduli. This may however not be the optimal path since it is expected that in such models the flux discretuum is also smaller.
 
  Second, we can choose fluxes implementing a hierarchy in the saxion values. If, for example, $s, u_1 \gg u_\lambda$ for $\lambda > 1$, we find ${\cal A}_i \gg 1$ for $i = 2,\ldots,m$. As a result, all $\nu_\lambda$ for $\lambda > 1$ are stabilized (supersymmetrically) at a higher scale. Only the lightest axion(s), in our case a linear combination of $\sigma$ and $\nu_1$, will remain relevant, experiencing an effective winding potential which follows from \eqref{eq:MultipleAxions}. It can be recast in the simpler form \eqref{eq:AxionPotential}, which we have studied in great detail.
  
  As a result, we expect that non-zero minima of the complex-structure $F$-term potential arise for appropriate choices of CY and flux. As discussed in the beginning of this section, we should be able to parametrically separate the potential difference between false and true vacuum, $\Delta V \sim e^K \epsilon^2 \gamma^2$, and the potential barrier $\sim e^K \epsilon^2$. In our regime of parametric control, $\epsilon \ll \abs{W_0}$, the uplift remains small compared to the depth $e^K|W_0|^2$ of the DGKT AdS vacuum. So, consistently with the no-go theorem of \cite{Hertzberg:2007wc,Flauger:2008ad}, dS is out of reach.
  
  However, due to the parametric separation between uplift and potential barrier, the brane tension $T$ between true and false vacuum may be too large for a bubble of true vacuum to nucleate, $T/M_\Pl > \sqrt{4/3} (\sqrt{\abs{V_t}}-\sqrt{\abs{V_f}})$ \cite{Cvetic:1992st, Narayan:2010em, Harlow:2010az}. The resulting non-SUSY AdS vacuum would be absolutely stable against the Coleman-de Luccia decay \cite{Coleman:1980aw} to the underlying SUSY AdS vacuum. If no other decay path exists, and we see no obvious candidate in the present setting, this would provide a counterexample to the non-SUSY AdS conjecture. Crucially, given the minimalist set of ingredients in our construction, it may actually be possible to study the type-IIA flux landscape in concrete models and establish, using e.g.~the technology developed in \cite{Demirtas:2019sip}, that the small uplifts we propose really exist.

 \section{Conclusion}\label{sec:Conclusion}
  
  We have presented a mechanism for metastable SUSY breaking in the landscape. Based on the winding scenario of \cite{Hebecker:2015rya}, we have described how the interplay of multiple periodic terms in the complex-structure super- and K\"ahler potential can lead to an $F$-term scalar potential with non-trivial local minima. Crucially, we have argued that the tuning power of the complex-structure landscape can be used to ensure that these minima are at parametrically small value of the potential, resulting in long-lived vacua. As applications, we discussed the uplift of LVS, DGKT and KKLT AdS vacua.

  KKLT is the most difficult case. Here, the requirement of a small perturbative superpotential $W_0$ inhibits the straightforward application of our simplest setup from Sect.~\ref{sec:MainSection}. The problem is that not only an axionic component Re$(\phi)$, but the full complex field $\phi$ remains light. While we described how our mechanism may still work, using a specific tuning of $W_0$ and of the additional periodic terms, we had to leave the explicit calculation of the minimum value and stability of the Im$(\phi)$-component for future work.
  
  For supersymmetric DGKT AdS vacua, our SUSY breaking mechanism appears to be very robust. Compared to KKLT, the construction requires less tuning. In fact, the DGKT setting naturally gives rise to a multi-axion potential which also comes with more tunable parameters. As a remark of caution, we note that the tuning power of the type-IIA complex structure landscape is less established than its type-IIB counterpart. (A brief discussion of how the necessary tuning could be implemented appears  below \eqref{eq:DGKTVacuum}.) Assuming that this concern can be dispelled, we may have found stable non-supersymmetric AdS solutions serving as counterexamples to the non-SUSY AdS conjecture~\cite{Ooguri:2016pdq}.
  
  Finally, we showed that the application of our winding uplift in the LVS context is straightforward. Of course, in this case one is working on the basis of a non-supersymmetric AdS solution, which may in itself already be in conflict with the non-SUSY AdS conjecture (cf.~the discussion in \cite{Conlon:2018vov}). If the LVS AdS vacuum exists, our uplift provides a new route to metastable dS vacua, challenging the de Sitter conjecture \cite{Ooguri:2018wrx}. The strong point of this uplift is its conceptual and technical simplicity, allowing in principle for the computerized search for a completely explicit example.  A summary of the main tuning requirements that have to be met appears at the end of Sect.~\ref{sec:AxionPotential}. We want to highlight the very recent work of \cite{Carta:2021sms} where the parameters $\alpha$ and $\epsilon$ of \eqref{eq:AxionPotential} were made explicit in a CICY setup. Unfortunately this CICY construction does not allow for an implementation of the LVS, so some more work needs to be done.
   
  Before closing, we recall that we started our work with a discussion of possible extended swampland conjectures against non-SUSY AdS. We do not want to repeat this discussion but only emphasize one possible landscape-skeptical scenario: If the LVS AdS vacuum and the KKLT uplift (by our or other methods) fail, then our only application is DGKT. Here, our uplift to non-SUSY AdS may work, but it appears difficult to raise the $F$-term scale far above the AdS scale. Such a situation would support a `Strong-SUSY-Breaking Conjecture', excluding non-SUSY AdS vacua (stable or metastable) with an $F$-term scale parametrically above the AdS scale. This conjecture is interesting as it rules out a type of cosmology which would otherwise be perfectly consistent with most of the swampland conjectures and with observers like us: Namely, a universe just like ours but where the late stages of cosmology are governed by a tiny negative cosmological constant. If one takes the landscape/multiverse view on fundamental physics seriously (see \cite{Hebecker:2020aqr} for a recent review), the existence or non-existence of such cosmologies in string theory represents an important question.

 \vspace{1cm}
  
 \phantomsection
 \subsection*{Acknowledgments} 
  We would like to thank Daniel Junghans for useful discussions. A.H. acknowledges an enjoyable conversation with Alexander Westphal about possible swampland conjectures against SUSY breaking which has inspired some of the ideas discussed in this paper. This work is supported by the Deutsche Forschungsgemeinschaft (DFG, German Research Foundation) under Germany's Excellence Strategy EXC 2181/1 - 390900948 (the Heidelberg STRUCTURES Excellence Cluster) and  the  Gra\-du\-ier\-ten\-kol\-leg `Particle  physics  beyond  the  Standard  Model' (GRK  1940).
 
 \vspace{1cm}

 \phantomsection
 \addcontentsline{toc}{section}{References}
 \bibliographystyle{JHEP}
 \bibliography{references}

\end{document}

%% file: figures/ZerosIntro.pdf_tex
\begingroup%
  \makeatletter%
  \providecommand\color[2][]{%
    \errmessage{(Inkscape) Color is used for the text in Inkscape, but the package 'color.sty' is not loaded}%
    \renewcommand\color[2][]{}%
  }%
  \providecommand\transparent[1]{%
    \errmessage{(Inkscape) Transparency is used (non-zero) for the text in Inkscape, but the package 'transparent.sty' is not loaded}%
    \renewcommand\transparent[1]{}%
  }%
  \providecommand\rotatebox[2]{#2}%
  \newcommand*\fsize{\dimexpr\f@size pt\relax}%
  \newcommand*\lineheight[1]{\fontsize{\fsize}{#1\fsize}\selectfont}%
  \ifx\svgwidth\undefined%
    \setlength{\unitlength}{569.22981262bp}%
    \ifx\svgscale\undefined%
      \relax%
    \else%
      \setlength{\unitlength}{\unitlength * \real{\svgscale}}%
    \fi%
  \else%
    \setlength{\unitlength}{\svgwidth}%
  \fi%
  \global\let\svgwidth\undefined%
  \global\let\svgscale\undefined%
  \makeatother%
  \begin{picture}(1,0.43715934)%
    \lineheight{1}%
    \setlength\tabcolsep{0pt}%
    \put(0,0){\includegraphics[width=\unitlength,page=1]{ZerosIntro.pdf}}%
    \put(-0.035,0.37){\color[rgb]{0,0,0}\makebox(0,0)[lt]{\footnotesize $V_\text{wall}$}}%
    \put(0.00,0.025){\color[rgb]{0,0,0}\makebox(0,0)[lt]{\footnotesize $-\pi/2$}}%
    \put(0.488,0.025){\color[rgb]{0,0,0}\makebox(0,0)[lt]{\footnotesize $0$}}%
    \put(0.90,0.025){\color[rgb]{0,0,0}\makebox(0,0)[lt]{\footnotesize $\pi/2$}}%
    \put(-0.05,0.075){\color[rgb]{0,0,0}\makebox(0,0)[lt]{\footnotesize $\Delta V$}}%
    \put(0.015,0.455){\color[rgb]{0,0,0}\makebox(0,0)[lt]{\footnotesize $V(\varphi)$}}%
    \put(0.995,0.052){\color[rgb]{0,0,0}\makebox(0,0)[lt]{\footnotesize $\varphi$}}%
  \end{picture}%
\endgroup%

%% file: figures/zeros.pdf_tex
\begingroup%
  \makeatletter%
  \providecommand\color[2][]{%
    \errmessage{(Inkscape) Color is used for the text in Inkscape, but the package 'color.sty' is not loaded}%
    \renewcommand\color[2][]{}%
  }%
  \providecommand\transparent[1]{%
    \errmessage{(Inkscape) Transparency is used (non-zero) for the text in Inkscape, but the package 'transparent.sty' is not loaded}%
    \renewcommand\transparent[1]{}%
  }%
  \providecommand\rotatebox[2]{#2}%
  \newcommand*\fsize{\dimexpr\f@size pt\relax}%
  \newcommand*\lineheight[1]{\fontsize{\fsize}{#1\fsize}\selectfont}%
  \ifx\svgwidth\undefined%
    \setlength{\unitlength}{1251.99993896bp}%
    \ifx\svgscale\undefined%
      \relax%
    \else%
      \setlength{\unitlength}{\unitlength * \real{\svgscale}}%
    \fi%
  \else%
    \setlength{\unitlength}{\svgwidth}%
  \fi%
  \global\let\svgwidth\undefined%
  \global\let\svgscale\undefined%
  \makeatother%
  \begin{picture}(1,0.30111822)%
    \lineheight{1}%
    \setlength\tabcolsep{0pt}%
    \put(0,0){\includegraphics[width=\unitlength,page=1]{zeros.pdf}}%
    \put(0.20,0.27){\color[rgb]{0,0,0}\makebox(0,0)[lt]{\footnotesize $F$}}%
    \put(0.20,0.13){\color[rgb]{0,0,0}\makebox(0,0)[lt]{\footnotesize $V$}}%
    \put(0.855,0.093){\color[rgb]{0,0,0}\makebox(0,0)[lt]{\footnotesize $V_0$}}%
    \put(0.60,0.093){\color[rgb]{0,0,0}\makebox(0,0)[lt]{\footnotesize $V_0$}}%
    \put(0.345,0.093){\color[rgb]{0,0,0}\makebox(0,0)[lt]{\footnotesize $V_0$}}%
    \put(0.09,0.093){\color[rgb]{0,0,0}\makebox(0,0)[lt]{\footnotesize $V_0$}}%
	
    \put(0.005,0.179){\color[rgb]{0,0,0}\makebox(0,0)[lt]{\footnotesize $-\frac{\pi}{2}$}}%
    \put(0.200,0.179){\color[rgb]{0,0,0}\makebox(0,0)[lt]{\footnotesize $\frac{\pi}{2}$}}%
    \put(0.260,0.179){\color[rgb]{0,0,0}\makebox(0,0)[lt]{\footnotesize $-\frac{\pi}{2}$}}%
    \put(0.455,0.179){\color[rgb]{0,0,0}\makebox(0,0)[lt]{\footnotesize $\frac{\pi}{2}$}}%
    \put(0.515,0.179){\color[rgb]{0,0,0}\makebox(0,0)[lt]{\footnotesize $-\frac{\pi}{2}$}}%
    \put(0.710,0.179){\color[rgb]{0,0,0}\makebox(0,0)[lt]{\footnotesize $\frac{\pi}{2}$}}%
    \put(0.770,0.179){\color[rgb]{0,0,0}\makebox(0,0)[lt]{\footnotesize $-\frac{\pi}{2}$}}%
    \put(0.965,0.179){\color[rgb]{0,0,0}\makebox(0,0)[lt]{\footnotesize $\frac{\pi}{2}$}}%
	
    \put(0.005,0.012){\color[rgb]{0,0,0}\makebox(0,0)[lt]{\footnotesize $-\frac{\pi}{2}$}}%
    \put(0.200,0.012){\color[rgb]{0,0,0}\makebox(0,0)[lt]{\footnotesize $\frac{\pi}{2}$}}%
    \put(0.260,0.012){\color[rgb]{0,0,0}\makebox(0,0)[lt]{\footnotesize $-\frac{\pi}{2}$}}%
    \put(0.455,0.012){\color[rgb]{0,0,0}\makebox(0,0)[lt]{\footnotesize $\frac{\pi}{2}$}}%
    \put(0.515,0.012){\color[rgb]{0,0,0}\makebox(0,0)[lt]{\footnotesize $-\frac{\pi}{2}$}}%
    \put(0.710,0.012){\color[rgb]{0,0,0}\makebox(0,0)[lt]{\footnotesize $\frac{\pi}{2}$}}%
    \put(0.770,0.012){\color[rgb]{0,0,0}\makebox(0,0)[lt]{\footnotesize $-\frac{\pi}{2}$}}%
    \put(0.965,0.012){\color[rgb]{0,0,0}\makebox(0,0)[lt]{\footnotesize $\frac{\pi}{2}$}}%
  \end{picture}%
\endgroup%